\documentclass[10pt]{spieman}  
\usepackage[]{graphicx}
\usepackage{setspace}
\usepackage{tocloft}

\usepackage{color}

\title{Four-wave mixing in a ring cavity}

\author{Eugeniy E. Mikhailov,\supscr{a} Jesse Evans,\supscr{a} Dmitry Budker,\supscr{b,c,d} Simon M. Rochester\supscr{b}, Irina Novikova\supscr{a}}

\affiliation{\supscrsm{a}Department of Physics, The College of William and Mary, Williamsburg, VA 23185, USA\\
\supscrsm{b}Rochester Scientific, LLC, El Cerrito, CA, 94530, USA\\
\supscrsm{c}Department of Physics, University of California, Berkeley, CA 94720, USA\\
\supscrsm{d}Helmholtz Institute Mainz, Johannes Gutenberg University, 55099 Mainz, Germany}

\cftpagenumbersoff{figure}
\cftpagenumbersoff{table}
\begin{document}
\maketitle

\begin{abstract}
We investigate  a  four-wave mixing process in an \emph{N}
interaction scheme in Rb vapor placed inside a low-finesse ring
cavity. We observe strong amplification and generation of a
probe signal, circulating in the cavity, in the presence of two
strong optical pump fields. We study the variations in probe
field gain and dispersion as functions of experimental
parameters with an eye on potential application of such a
system for enhanced rotation measurements. A density-matrix
calculation is performed to model the system, and the
theoretical results are compared to those of the experiment.
\end{abstract}


{\noindent \footnotesize{\bf Address all correspondence to}:
Eugeniy E.  Mikhailov, The College of William and Mary, Department of Physic,  Williamsburg,
P.O. Box 8795, VA 23187, USA;
Tel: +1 757-221-3571; Fax: +1 757-221-3540; E-mail: \linkable{eemikh@wm.edu}
}

\begin{spacing}{2}   

\section{Introduction}
\label{sect:intro}  

Recent theoretical proposals, predicting a significant improvements in the laser gyro sensitivity with the use of a ``white-light'' cavity~\cite{shahriarPRA07}, have motivated a number of theoretical and experimental studies of negative dispersion in various  systems~\cite{shahriarPRL07,shahriarOC08,ISI:000274791200119,Salit:11,ISI:000314911400027}.
A promising approach relies on manipulation of optical dispersion via coherent interactions of light with resonant atomic media. Processes like electromagnetically induced transparency (EIT), stimulated Raman scattering (SRS), four-wave mixing (FWM), etc., are known to enable group index $n_g$ variation from subluminal (``slow light'', $n_g\gg 1$) to superluminal  (``fast light'', $n_g<1$)~\cite{Boyd2002}.
Demonstration of a white-light laser gyroscope requires  a gain medium with
group index tunable around $n_g=0$. Several experimental groups have pursued
various approaches for the realization of such
conditions~\cite{shahriarPRL07,ShahriarOE10,YifuZhuPRA10,Salit:11,Kotlicki:12}.

\begin{figure}
   \begin{center}
   \begin{tabular}{c}
   \includegraphics[width=0.4\columnwidth]{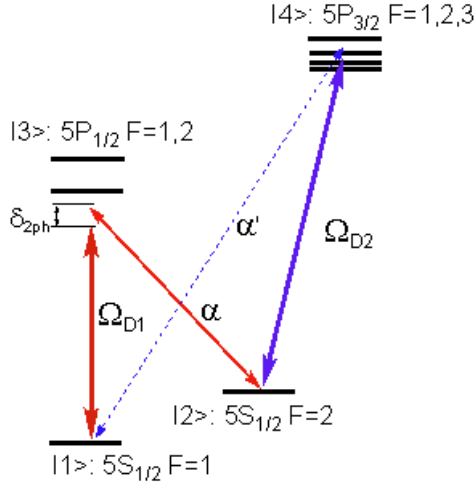}
   \end{tabular}
   \end{center}
   \caption
   { \label{fig:levels} A simplified \emph{N}-type interaction
scheme in a four-level system, and its practical realization in
${}^{87}$Rb. The two pump fields are labeled $\Omega_{D1}$ and
$\Omega_{D2}$, the probe field $\alpha$ forms a
$\Lambda$ link with the $\Omega_{D1}$ pump field. The fourth optical field $\alpha^\prime$ is
generated to complete the four-photon four-wave mixing process
(not monitored in the experiment). The two-photon detuning $\delta_\mathrm{2ph}$ is the frequency difference between the $D_1$ pump field $\Omega_1$ and the probe field $\alpha$ minus the frequency of the $|2\rangle-|1\rangle$ transition. }
   \end{figure}

Recently our groups demonstrated that the fast light regime with
amplification can be achieved in a four-level \emph{N}
interaction scheme, shown in
Fig.~\ref{fig:levels}~\cite{Phillips2013,SPIEProc2013}. In such
a configuration the two optical fields---the strong pump field
$\Omega_{D1}$ and weak signal field $\alpha$---form a regular
$\Lambda$ system exhibiting EIT and slow
light~\cite{lukin03rmp}. The four-level \emph{N}-scheme is then
completed with the introduction of the second strong pump field
$\Omega_{D2}$. A strong amplification of the probe field is
possible in the case that the selection rules allow the
four-wave mixing
process~\cite{lukinPRLFWM99,Mikhailov2002,phillipsJMO09,LukinPhysRevA.79.013806,PhillipsPhysRevA.83.063823}.
At the same time, the additional Rabi splitting introduced by
the second strong optical field provides an extra control
mechanism for the probe field
dispersion\cite{FleischhakerPRA08,AbiSalloum2009,LettFWM12}.
The possibility of probe field propagation with no optical
losses or with gain, with a smooth transition between slow- and
fast-light regimes by varying the strength of one of the pump
fields ($\Omega_{D2}$) was theoretically
demonstrated~\cite{Phillips2013,SPIEProc2013}, making it an
ideal test bed for explorations of the dispersion effects in a
ring cavity laser for gyroscope applications.

In this paper we study the optical characteristics of the \emph{N}-system
(Fig.~\ref{fig:levels}) when the weak probe field $\alpha$ is coupled
to a low-finesse ring cavity. This is a step towards experimentally testing the concept of a white-light-cavity gyroscope. We observed strong amplification of the
circulating probe optical field under the FWM conditions. Moreover, even in the
absence of the seeded input, an optical field at the probe frequency was
generated inside the cavity in the presence of two pump fields, with a
characteristic laser threshold behavior with respect to the power of either pump. We
also explored the group delay between input and output amplitude-modulated
probe fields, and observed smooth tuning from positive delay (slow light) to
negative delay (fast light) depending on the probe two-photon detuning.

\section{Experimental apparatus}

   \begin{figure}
   \begin{center}
   \begin{tabular}{c}
   \includegraphics[width=0.8\columnwidth]{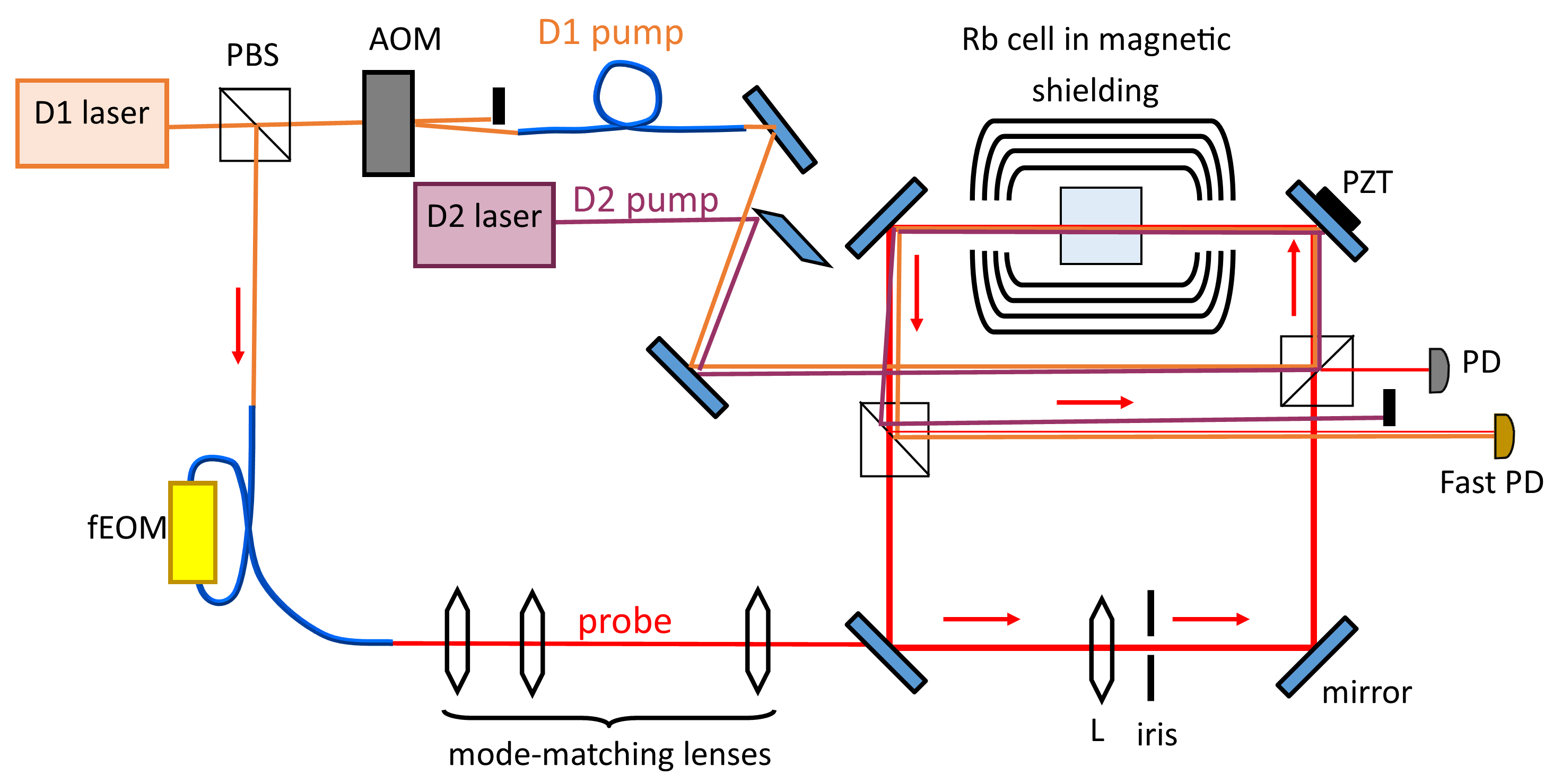}
   \end{tabular}
   \end{center}
   \caption
   { \label{fig:setup}
A schematic of the experimental setup. See text for abbreviations. }
   \end{figure}

A simplified schematic of the experimental setup is shown in
Fig.~\ref{fig:setup}. The current version of the experiment
uses two external cavity diode lasers, each of which are
equipped with a saturation-absorption-spectroscopy frequency
reference, not shown in the figure.
One laser, tuned to the $D_2$ optical transition of
${}^{87}$Rb, serves as the $\Omega_{D2}$ pump field, and the
second laser, operating at the ${}^{87}$Rb $D_1$ line
($795$~nm) is used to generate both $\Omega_{D1}$ pump and
$\alpha$ probe optical fields. For that purpose the laser
output is split with a polarizing beam splitter (PBS). The
transmitted beam is coupled into an single-mode optical fiber,
and phase-modulated using a fiber-coupled electro-optical
modulator (fEOM) operating at the microwave frequency of
$6.91468$~GHz$+\delta_\mathrm{2ph}$. The power of the
modulation was chosen such that the amplitude of the
unmodulated carrier field was suppressed to approximately
$20\%$ of its original value, and approximately 40\% of the
power was transferred to each of the first modulation
sidebands. The output of the fiber modulator is coupled into
the ring-cavity, with additional lenses to ensure maximum
coupling into the fundamental TEM${}_{00}$ spatial cavity mode.

The two pump fields are combined using a sharp-edge mirror at a
small angle of approximately $7$~mrad and enter the cavity
through the input polarizing beam-splitter. They are
linearly-polarized and perpendicular to the intra-cavity probe
field polarization. The direct output of the $780$~nm laser
(properly attenuated and linearly polarized) is used for the
$D_2$ pump. The laser beam at the cell is oval-shaped, with
minimum and maximum diameters of $2$~mm and $3$~mm. The
frequency of the $\Omega_{D1}$ pump is shifted down by $80$~MHz
using an acousto-optical modulator (AOM) with respect to the
$795$~nm laser output to ensure that the frequency difference
between this optical field and the probe field ($\alpha$)
$6.83468$~GHz$+\delta_\mathrm{2ph}$ is close to the ${}^{87}$Rb
ground state hyperfine splitting. This additional shift in
optical frequency of the $D_1$ pump helps to eliminate a
parasitic interference between this field and the carrier
frequency component of the fEOM output (of the original laser
wavelength), which partially leaks into the ring cavity. After
passing through a single-mode optical fiber, the $D_1$ pump
field is collimated into a Gaussian beam with a waist of
$2.5$~mm FWHM.

The ring cavity consists of four flat mirrors ($99.5$~\%
reflection), arranged in a square configuration, and a convex
lens (focal length $30$~cm). The total length of the cavity is
$77$~cm. One of the mirrors is mounted on a piezo-electric
device (PZT), which allows sweeping the cavity length to
observe its resonances or locking the cavity resonance
frequency to the frequency of the circulating optical field
using a feedback loop.

For the current experiments we used a cylindrical Pyrex Rb
vapor cell (diameter 22~mm; length 25~mm), containing
isotopically enriched ${}^{87}$Rb and 5~torr of Ne buffer gas,
placed inside a three-layer magnetic shielding to mitigate the
effects of any stray magnetic field. The cell is heated to
$90^\circ$, and the value of the column density $1.2\times
10^{12}~\mathrm{cm}^{-2}$ was extracted  by fitting the
weak-probe absorption curve.
To realize the four-wave mixing configuration for the probe
optical field inside the ring cavity, we have also added two
polarizing beam splitters inside the cavity such that the
circulating probe field is transmitted, but the two pump
optical fields are reflected to overlap with the probe field on
the PBS, and then reflected off at the output PBS. After
introduction of these optical elements the cavity finesse is
approximately $\mathcal{F}=18$ for light frequencies not on
resonance with the Rb atoms, which means that the power of the
circulating optical field reduced to approximately
$\gamma\simeq 0.70$ of its initial value after one round-trip
inside the cavity~\cite{Siegman1986}:
\begin{equation}\label{finess}
    \mathcal{F}=\frac{\pi\sqrt[4]{\gamma}}{1-\sqrt{\gamma}}.
\end{equation}
The free spectral range of the cavity (FSR) was measured to be
$347$~MHz. Since the different frequencies outputted by the EOM
are not simultaneously resonant with the cavity, we are able to
couple in only the probe field, without need for additional
spectral filtering.

We used a small residual reflection of the probe optical field
at the input PBS to monitor its power. To measure its spectral
properties we directed the $D_1$ pump field and the probe
field, reflected off the output cavity PBS, into a fast
photodetector, and monitored their beatnote amplitude and
frequency using a spectrum analyzer.

\section{Experimental measurements of the probe field amplification and generation}

\begin{figure}
   \begin{center}
   \begin{tabular}{c}
   \includegraphics[width=0.8\columnwidth]{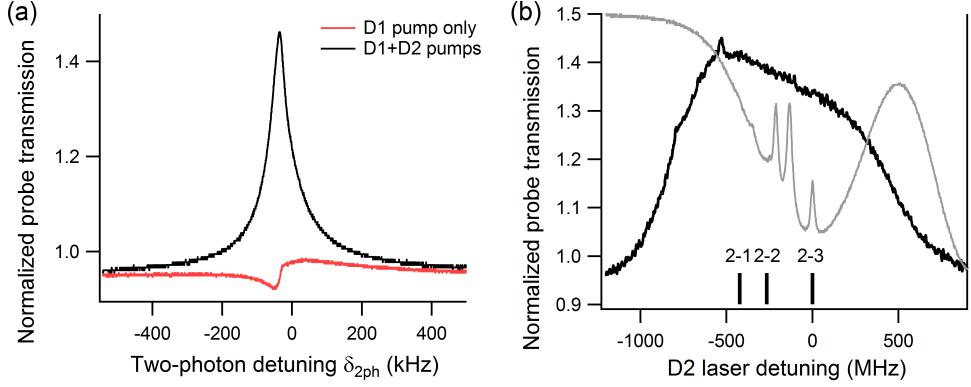}
   \end{tabular}
   \end{center}
   \caption
   { \label{fig:singlpassgain}
\emph{(a)} Examples of the single-pass probe-field transmission through the Rb cell as a function of two-photon detuning $\delta_{2ph}$ in the presence of only $\Omega_{D1}$ optical field ($\Lambda$-scheme) and of both $\Omega_{D1}$ and $\Omega_{D2}$ optical fields (\emph{N}-scheme). For these measurements the $D_1$ laser is tuned $\approx 400$~MHz to the red from the $5S_{1/2}F=1\rightarrow 5P_{1/2}F'=1$ transition, and the $D_2$ laser is tuned near the $5S_{1/2}F=2\rightarrow 5P_{3/2}F'=1$ transition.
\emph{(b)} Single-pass amplification of the probe field (black line) as a function of the $D_2$ laser optical detuning (zero detuning corresponds to $F=2\rightarrow F'=3$ transition). The $D_1$ laser detuning is the same as in \emph{(a)}, and the two-photon detuning is zero. The saturation-absorption spectrum of a reference cell (grey line), as well as the positions of $5S_{1/2}F=2\rightarrow 5P_{3/2}F'$ resonances are also shown. The powers of the $D_1$ and $D_2$ pump lasers are 0.7~mW  and 5.4~mW correspondingly.
}
   \end{figure}

Before placing the atomic medium in the ring cavity we
confirmed that it is possible to  achieve probe field
amplification in our realization of the \emph{N} interaction
scheme. In particular, if only the $D_1$ pump field was
present, completing a single $\Lambda$ link, we observe the
usual EIT-like variation in the probe transmission with the
two-photon detuning between the probe and the pump fields.
Fig.~\ref{fig:singlpassgain}(a) demonstrates the two-photon
resonance for the $\Omega_{D1}$ laser detuning at the slope of
the Doppler-broadened optical transition, chosen to match the
experimental conditions in which we observed maximum FWM gain
later. Because of this relatively large laser detuning, the
shape of the two-photon resonance was
asymmetric~\cite{mikhailov04}. We have maintained this detuning
value of the $D_1$ laser from the optical transition frequency
for all reported measurements. However, it is possible that the
value of the laser detuning corresponding to the maximum gain
depends on the temperature of the cell, i.e., the resonant
absorption can overcome gain produced in the four-wave mixing
process.  Dependence of FWM on atomic density will be a subject
of further studies.

In the presence of the second pump field, forming the
\emph{N}-scheme, the probe field exhibited a strong symmetric
amplification peak, with maximum gain of $\approx 1.5$. We also
measured the variation of the probe gain with the $D_2$ laser
detuning, as shown in Fig.~\ref{fig:singlpassgain}(b). While
amplification was observed in a relatively wide range of laser
frequencies, the highest gain occurred around
$5S_{1/2}F=2\rightarrow 5P_{3/2}F'=1,2$ transitions, for which
the selection rules allow the four-wave mixing process.

\begin{figure}
   \begin{center}
   \begin{tabular}{c}
   \includegraphics[width=0.8\columnwidth]{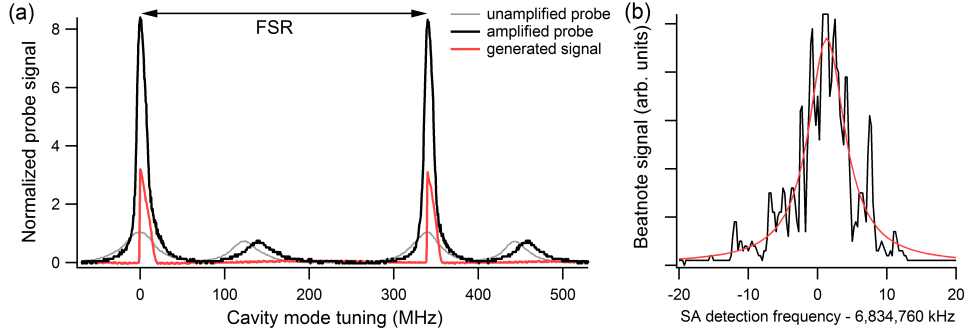}
   \end{tabular}
   \end{center}
   \caption
   { \label{fig:cavitytransm}
\emph{(a)} Cavity transmission for seeded input probe field with amplification (both pump fields are on) and without amplification ($D_1$ pump is off), as well as spontaneous generation of the probe field (no probe seeding, both pump are on). The additional smaller peaks are the off-resonant ``+1''  fEOM modulation sideband.
\emph{(b)} The spectrum of a beatnote between the generated probe field and the $D_1$ pump field. Both resolution and video bandwidth of the spectrum analyzer were set on $1$~kHz.
The $D_1$ laser is tuned approx. $400$~MHz to the red from the $5S_{1/2}F=1\rightarrow 5P_{1/2}F'=1$ transition, the $D_2$ laser is tuned near the $5S_{1/2}F=2\rightarrow 5P_{3/2}F'=2$ transition, and the two-photon detuning is $-30$~kHz. The powers of the $D_1$ and $D_2$ pump lasers are 0.7~mW  and 5.4~mW correspondingly.}
   \end{figure}

We observed even stronger amplification when the probe field
circulated inside the ring cavity.
Fig.~\ref{fig:cavitytransm}(a) shows that when both pump fields
were tuned to the four-wave mixing gain conditions, the probe
field output increased dramatically.  While we expect such
amplification to be accompanied by the generation of the fourth
optical field on the $|4\rangle\rightarrow|1\rangle$ optical
transition (as we observed in previous
studies~\cite{SPIEProc2013}), we were not able to directly
detect it here, as the experimental arrangement does not favor
its amplification. Indeed, this field has to propagate along
the $D_2$ pump field to satisfy the phase-matching conditions,
and thus it cannot circulate inside the ring cavity due to the
small misalignment of the $D_2$ pump field.

When the input probe field blocked, so that only the two pump
fields interacted with the atoms, we observed the generation of
an optical field at the probe frequency, circulating inside the
cavity. The analysis of the beatnote between this new generated
field and the $D_1$ pump field indicated that these two fields
are phase-coherent. An example of such a beatnote shown in
Fig.~\ref{fig:cavitytransm}(b) gives a rough width estimate of
$6$~kHz (FWHM), determined predominantly by the acoustic
instabilities of the cavity. The instantaneous width seems to
be limited only by the spectrum-analyzer resolution.

\begin{figure}
   \begin{center}
   \begin{tabular}{c}
   \includegraphics[width=0.8\columnwidth]{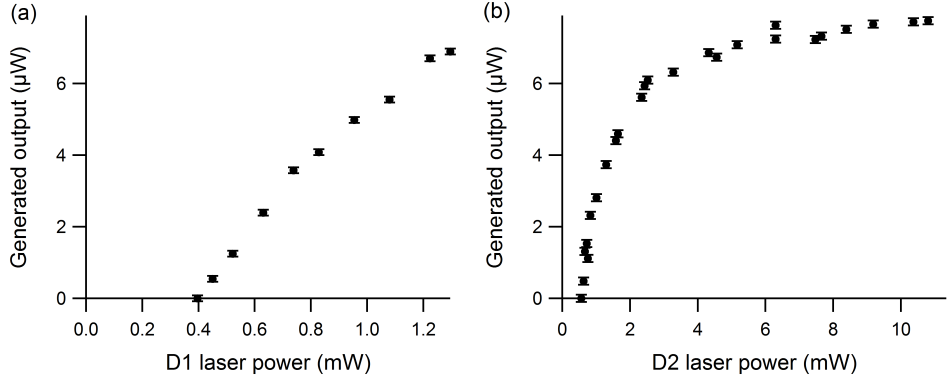}
   \end{tabular}
   \end{center}
   \caption
   { \label{fig:thresholds}
The generated probe power as a function of \emph{(a)} $D_1$ laser power ($D_2$ power maintained at 5.4~mW) and of \emph{(b)} $D_2$ laser power ($D_1$ power maintained at 1.3~mW). The laser detunings are the same as in Fig.~\ref{fig:cavitytransm}. }
   \end{figure}

The amplitude of the generated field displayed a typical
threshold behavior as a function of the power of either of the
pump beams, as shown in Fig.~\ref{fig:thresholds}. The values
of the thresholds also depended  on the laser detunings and
relative alignment of the pump fields. The fundamental cavity
mode had the lowest threshold values, which is not surprising,
since this was the mode the seeded probe field is coupled to,
and thus the pump laser parameters are optimized to maximize
the gain in that mode. However, at higher pump powers, higher
spatial modes were generated as well.  The threshold values for
those modes were typically twice as high as that for the
fundamental mode, but they depended strongly on the cavity
alignment and pump laser detunings. After the initial fast
increase of the generated field amplitude above the threshold,
we observed a clear saturation, for which further increase in
$D_2$ laser power did not produce significant growth in the
probe field power.

\section{Density-matrix simulation of the probe field propagation}


\label{sec:calculation}

A density-matrix calculation was performed to model the
propagation of the pump and probe fields through the atomic
medium, both for the case of a single pass of the probe light
through the medium, and the case in which the medium is placed
in a ring cavity. The atoms were modeled as a four-level system
as shown in Fig.\ \ref{fig:levels}. In analogy to the Rb
system, we refer to the pump fields as the $D_1$ and $D_2$
pumps, even though the intensities and detunings of these
fields were adjusted to account for the differences between
$^{87}$Rb and the simplified model system. The semiclassical
density-matrix evolution equations, written as a function of
distance along the light propagation direction, take into
account the optical fields, spontaneous decay from the upper
states, transit relaxation, and pressure broadening. In order
to take into account Doppler broadening, the longitudinal
velocity distribution is divided into velocity groups, with a
set of evolution equations written for each velocity group.
These equations are coupled to the one-dimensional wave
equation for the four optical fields, with the coupling for a
particular optical frequency proportional to the total atomic
polarization at that frequency, obtained by summing over all
velocity groups.

\begin{figure}[h]
\centering\includegraphics{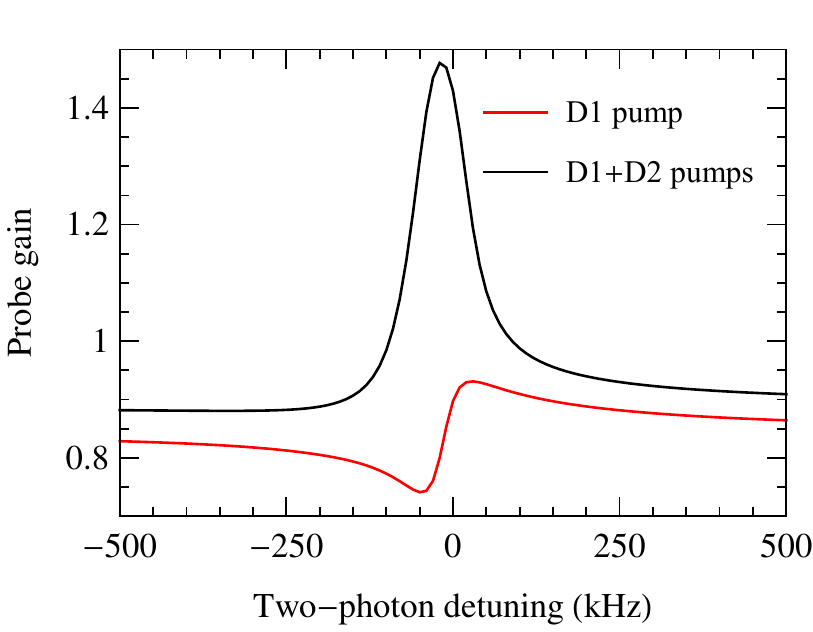}
\caption{Single-pass probe gain as a function of two-photon detuning in the presence of only the $D_1$ pump field and both the $D_1$ and $D_2$ pump fields.}\label{OnePassGainTheory}
\end{figure}

In the steady state, the equations reduce to a set of
differential algebraic equations in the spatial variable that
can be solved as an initial-value problem to find the change in
optical fields upon a single pass through the medium. Figure
\ref{OnePassGainTheory} shows the single-pass probe gain as a
function of two-photon detuning with D1 pump only and with both
D1 and D2 pump, for comparison with the experimental results of
Fig.\ \ref{fig:singlpassgain}. The following input parameters
were chosen to approximately match the parameters for $^{87}$Rb
and the experimental setup: excited state natural widths
$2\pi\times6$~MHz, pressure broadening $50$~MHz, Doppler width
330~MHz, optical fields' wavelengths 795~nm (for the $D_1$
optical transition) and 780~nm (for the $D_2$ optical
transition), and optical path length through the atomic medium
2.5~cm. The remaining experimental parameters were adjusted by
hand to improve the agreement with the experimental
measurements shown in Fig.~\ref{fig:singlpassgain}: transit
relaxation rate $2\pi\times20$~kHz, D1 pump detuning -800~MHz,
D2 pump detuning 0~MHz, D1 pump intensity 3~mW/cm$^2$, D2 pump
intensity 3~mW/cm$^2$, probe intensity 0.5~mW/cm$^2$, and
atomic density $9.6\times10^{10}$~cm$^{-3}$ (0.2 times the
experimental value; this scaling was necessary because the
four-level model system has intrinsically larger optical
coupling constants than Rb). These values were used for all of
the following results except where noted.

The propagation dynamics of the probe field inside the ring
cavity is characterized by two main parameters: the single-pass
complex gain through the atomic medium $g_{\mathrm{sp}}$, which
accounts for the effect of the interactions with the atomic
medium, and was calculated according to the method described
above; and the complex gain for one round trip of the probe
light through the empty cavity $g_{\mathrm{ec}}$ that describes
all intracavity losses and the accumulated optical phase shift.
(We use the term ``gain'' to describe $g_{\mathrm{ec}}$, even
though its absolute value is always less than unity.) The
generated probe amplitude and frequency are then obtained by
numerically solving for the condition of stable single-mode
oscillation, namely, that the total complex gain for one
round-trip inside the cavity is equal to unity:
$g_{\mathrm{sp}}\,g_{\mathrm{ec}}=1$.

\begin{figure}[h]
\centering
\includegraphics{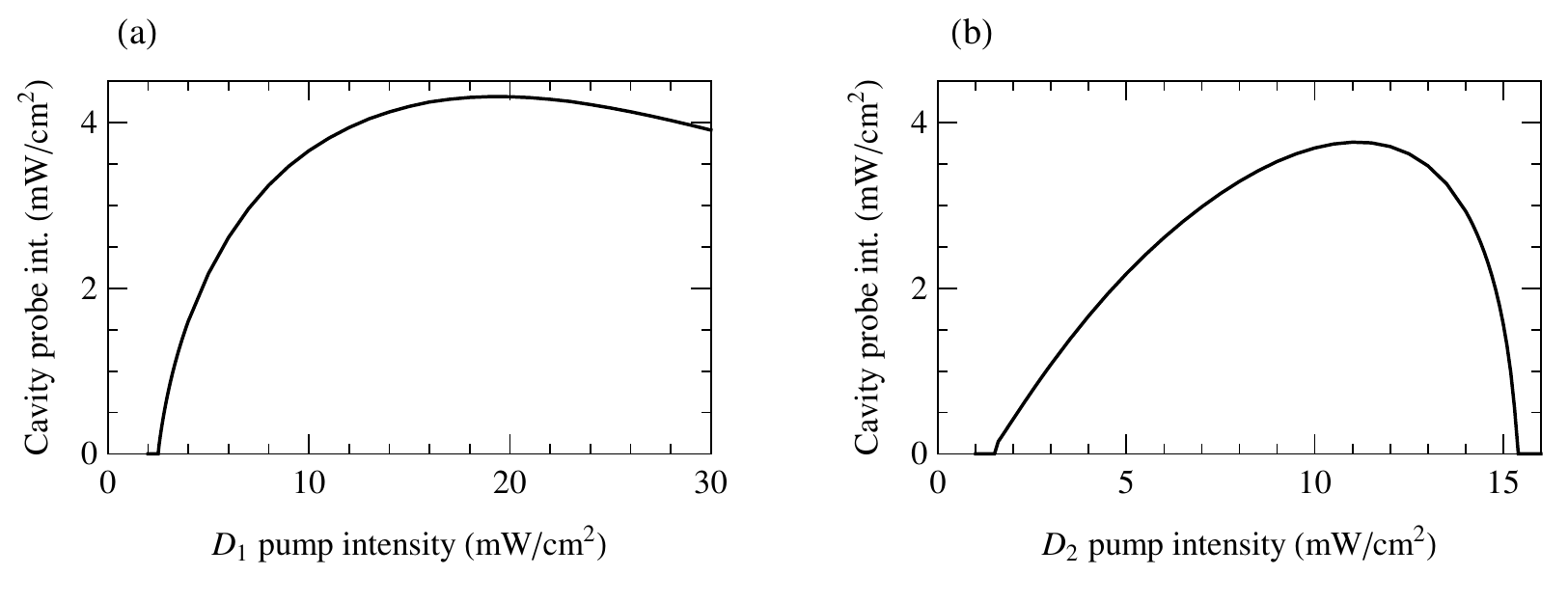}
\caption{Generated probe intensity (inside the cavity) as a
function of (a) D1 pump intensity and (b) D2 pump
intensity.}\label{GenProbeVsPumpsTheory}
\end{figure}
\begin{figure}[h]
\centering
\includegraphics{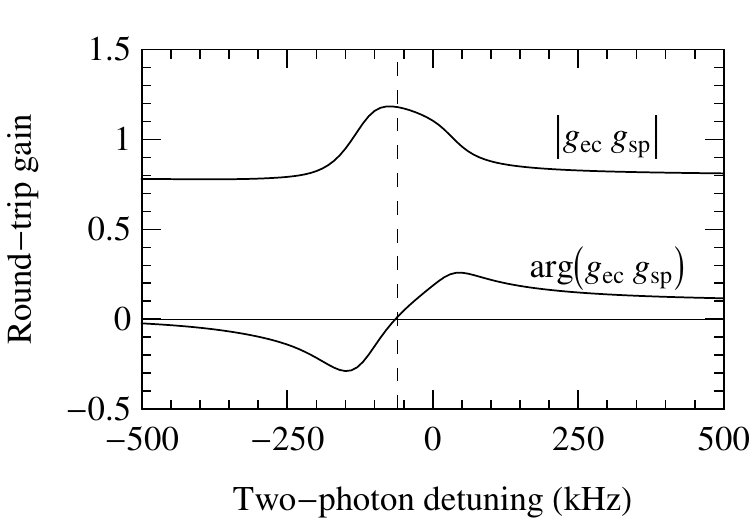}
\caption{Magnitude and argument of the complex round-trip gain
$g_{\mathrm{sp}}\,g_{\mathrm{ec}}$. The cavity has been tuned
so that the argument crosses zero within the gain
feature.}\label{RoundTripGainTheory}
\end{figure}

Figure \ref{GenProbeVsPumpsTheory} shows the circulating
generated probe intensity in the cavity as a function of D1 and
D2 pump intensities. The parameters are as given above, except
that the D2 pump intensity is fixed at 6~mW/cm$^2$ in Fig.\
\ref{GenProbeVsPumpsTheory}(a) and the the D1 pump intensity is
fixed at 6~mW/cm$^2$ in Fig.\ \ref{GenProbeVsPumpsTheory}(b).
The magnitude of the round-trip empty cavity gain
$|g_{\mathrm{ec}}|=\sqrt{\gamma}=0.84$ was determined from the
experimental finesse measurements in Eq.~\ref{finess}. In order
to allow lasing of the probe field, the complex phase of
$g_{\mathrm{ec}}$ was adjusted (corresponding to tuning the
cavity resonance frequency) so that the total round-trip
complex phase shift $\arg(g_{\mathrm{sp}}\,g_{\mathrm{ec}})$
crosses zero within the gain feature (Fig.\
\ref{RoundTripGainTheory}).

The calculated dependences of the generated probe intensity on
the intensities of each pump optical field (as the other pump
intensity is held fixed), shown in Fig.\
\ref{GenProbeVsPumpsTheory}, demonstrate clear generation
thresholds followed by a rapid increase in the generated probe
power, similar to the experimental observations. For higher
pump powers, however, the probe intensity levels off and then
falls. The beginning of this saturation was also observed in
Fig.~\ref{fig:thresholds}(b) as a function of $D_2$ pump power,
while the limited available power in the $D_1$ laser prevented
us from reaching this regime in Fig.~\ref{fig:thresholds}(a).
However, neither laser was powerful enough to observe the
reduction of the generated probe power in the limit of high
powers (although such fall-off was observed experimentally in
case of different laser detunings).

The calculation indicates that this behavior is a result of the
saturation and decrease of the single-pass gain
$g_{\mathrm{sp}}$ as a function of pump intensity. Even though
the magnitude of the single-pass gain does not fall below
unity, it does eventually fall below the threshold needed to
overcome the cavity losses and allow oscillation. The
saturation effect does not occur if both of the pump
intensities are increased simultaneously---evidently the
correct balance of pump intensities is required for optimal
probe gain in the medium.

\section{Delay/advance measurement}

   \begin{figure}
   \begin{center}
   \begin{tabular}{c}
   \includegraphics[width=0.6\columnwidth]{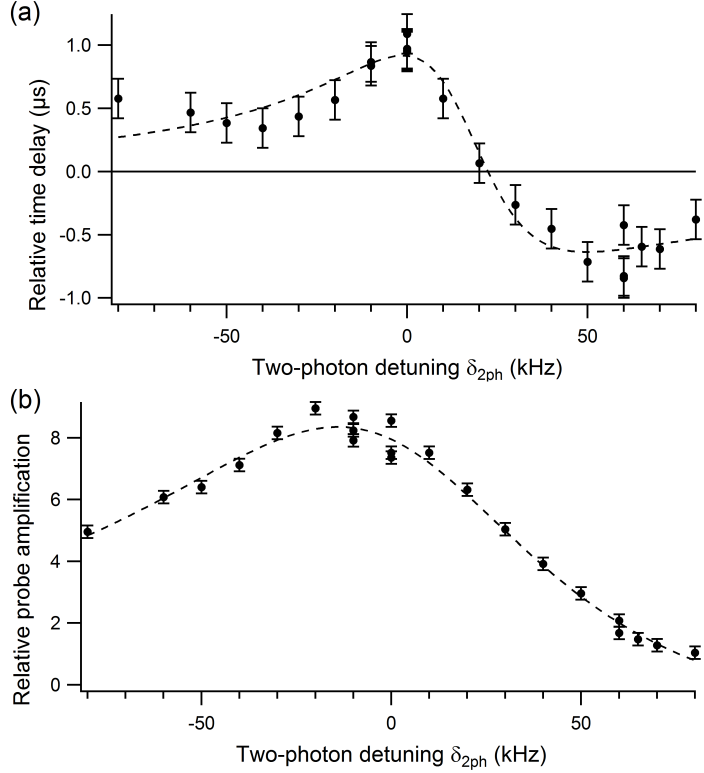}
   \end{tabular}
   \end{center}
   \caption
   { \label{fig:timedelay}
Relative time difference between the output and input sinusoidal amplitude modulation in the seeded probe signal [\emph({a})] and relative amplification of the cw probe signal [\emph({b})] as functions of the two-photon detuning. The dashed lines are to guide the eye. The powers of the pump lasers are 0.4~mW ($D_1$) and 4.0~mW ($D_2$), and the laser detunings are the same as in Fig.~\ref{fig:cavitytransm}. }
   \end{figure}

Next, we  evaluated the dispersive properties of the system by
sending an amplitude-modulated probe field and measuring its
relative time shift after the cavity. For these measurements we
actively locked the cavity on a probe-transmission resonance,
and in addition introduced a $10\%$ sinusoidal amplitude
modulation at frequency $f_{am}$. We have verified that with
careful choice of modulation parameters this modulation does
not affect the cavity lock. For these measurements we chose the
powers of the two pump fields such that only amplification and no
generation occurs in the cavity. The output probe signal maintained
the sinusoidal shape, and thus we determined the relative time
shift from changes in phase fitting parameter. The reference
phase is measured when both pump fields are blocked, and the
probe field is far-detuned from the atomic resonance.
Fig.~\ref{fig:timedelay} shows the variation of the observed time
shifts as we tuned the probe two-photon detuning through the
gain profile. Near the maximum we observed a relative delay.
However, for increasing negative detunings this delay smoothly
changed into advancement, indicating superluminal propagation.

\section{Cavity pulling}

\begin{figure}
   \begin{center}
   \begin{tabular}{c}
   \includegraphics[width=0.5\columnwidth]{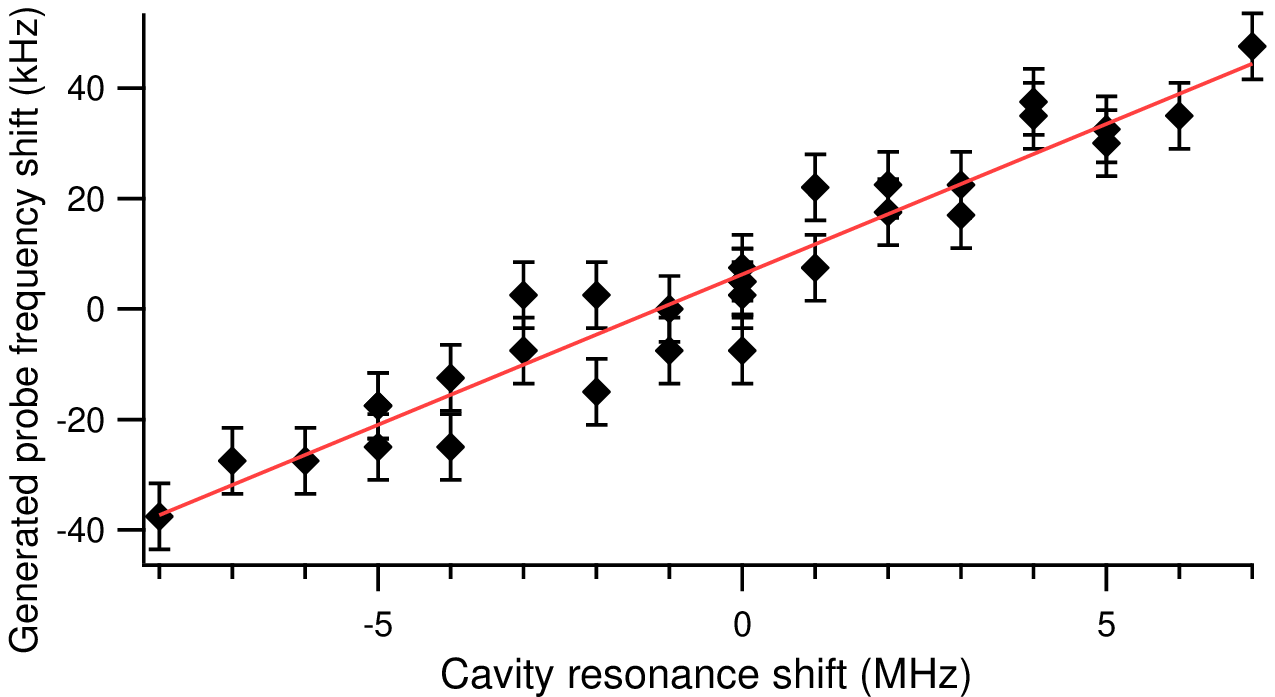}
   \includegraphics[width=0.5\columnwidth]{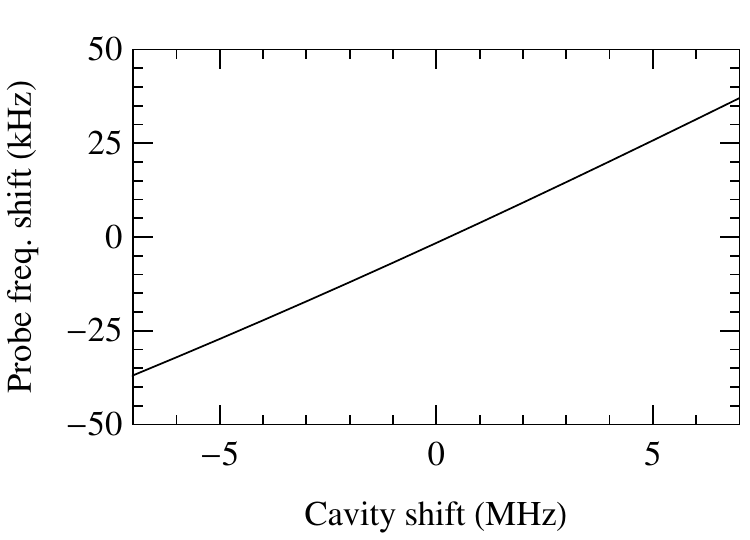}
   \end{tabular}
   \end{center}
   \caption
   { \label{fig:cavitypulling}
\emph{Left}: change in the generated field frequency as function of the cavity resonance tuning. The powers of the pump lasers are 0.6~mW ($D_1$) and 8.2~mW ($D_2$), and the laser detunings are the same as in Fig.~\ref{fig:cavitytransm}. \emph{Right}: corresponding numerical simulations, with $D_1$ and $D_2$ pump intensities both set to 6~mW/cm$^2$.  }
   \end{figure}

It is important to point out that the spectral bandwidth of the
FWM gain ($\le 100$~kHz) is significantly narrower than the
cavity transmission resonance ($\approx 20$~MHz), and thus our
current experiment cannot be considered a realization of the
original theoretical proposal~\cite{shahriarPRA07}. Under such
conditions we expect that the frequency and the width of the
probe field, generated inside a cavity, will be dominated by
the spectral properties of the $D_1$ laser field. Nevertheless,
we observed  that the change in the cavity length  produced a
measurable effect on the generated probe frequency, as shown in
Fig.~\ref{fig:cavitypulling}.

For these measurements we have injected into a cavity a ``+1''
modulation sideband detuned by approximately $150$~MHz such
that its maximum transmission occurred at the same cavity
length as the probe field generation. Since this field was
sufficiently weak and detuned far from any two-photon
resonances, its presence in the cavity did not affect FWM
process. At the same time, locking the cavity resonance to the
transmission peak of this seeded field allowed us to change the
cavity length in highly controllable manner, and observe a
systematic frequency shift of the generated probe field as the
cavity was swept through the transmission resonance. A
numerical calculation of the cavity pulling was also performed,
using the techniques described in Sec.~\ref{sec:calculation},
and it is in good agreement with the experimental results, as
seen in Fig.~\ref{fig:cavitypulling}(b).

\section{Conclusions}

In conclusion, we have analyzed the propagation of a weak resonant probe
through a medium of four-level atoms in an \emph{N}-scheme with allowed
four-wave mixing generation, and found it to be a promising
candidate for the realization of tunable ``slow-to-fast'' light
with no absorption. This is particularly interesting
for the experimental investigation of potential techniques for
the enhancement of optical-gyroscope performance.

\section{Acknowledgments}

The authors thank Frank A. Narducci and Jon Davis for useful discussions, and Gleb Romanov for helping with the experiment. This
research was supported by the Naval Air Warfare Center STTR program, contract N68335-13-C-0227.



\begin{thebibliography}{10}

\bibitem{shahriarPRA07}
M.~S. Shahriar, G.~S. Pati, R.~Tripathi, V.~Gopal, M.~Messall, and K.~Salit,
  ``Ultrahigh enhancement in absolute and relative rotation sensing using fast
  and slow light,'' {\em Phys. Rev. A} {\bf 75}(5), 053807  (2007).

\bibitem{shahriarPRL07}
G.~S. Pati, M.~Salit, K.~Salit, and M.~S. Shahriar, ``{Demonstration of a
  tunable-bandwidth white-light interferometer using anomalous dispersion in
  atomic vapor},'' {\em Phys. Rev. Lett.} {\bf {99}}  ({2007}).

\bibitem{shahriarOC08}
G.~S. Pati, M.~Salit, K.~Salit, and M.~S. Shahriar, ``Demonstration of
  displacement-measurement-sensitivity proportional to inverse group index of
  intra-cavity medium in a ring resonator,'' {\em Opt. Commun.} {\bf 281}(19),
  4931--4935  (2008).

\bibitem{ISI:000274791200119}
C.~Ciminelli, C.~E. Campanella, F.~Dell'Olio, and M.~N. Armenise, ``{Fast light
  generation through velocity manipulation in two vertically-stacked ring
  resonators},'' {\em Opt. Express} {\bf {18}}, {2973--2986}  ({2010}).

\bibitem{Salit:11}
M.~Salit, K.~Salit, and P.~Bauhahn, ``Prospects for enhancement of ring laser
  gyroscopes using gaseous media,'' {\em Opt. Express} {\bf 19}, 25312--25319
  (2011).

\bibitem{ISI:000314911400027}
O.~Kotlicki, J.~Scheuer, and M.~S. Shahriar, ``{Theoretical study on Brillouin
  fiber laser sensor based on white light cavity},'' {\em Opt. Express} {\bf
  {20}}, {28234--28248}  ({2012}).

\bibitem{Boyd2002}
R.~W. Boyd and D.~J. Gauthier, ``Chapter 6 ``slow'' and ``fast'' light,'' in
  {\em Progress in Optics},  E.~Wolf, Ed., {\em Progress in Optics} {\bf 43},
  497 -- 530, Elsevier  (2002).

\bibitem{ShahriarOE10}
H.~N. Yum, M.~Salit, J.~Yablon, K.~Salit, Y.~Wang, and M.~S. Shahriar,
  ``Superluminal ring laser for hypersensitive sensing,'' {\em Opt. Express}
  {\bf 18}, 17658--17665  (2010).

\bibitem{YifuZhuPRA10}
J.~Zhang, X.~Wei, G.~Hernandez, and Y.~Zhu, ``White-light cavity based on
  coherent raman scattering via normal modes of a coupled cavity-and-atom
  system,'' {\em Phys. Rev. A} {\bf 81}, 033804  (2010).

\bibitem{Kotlicki:12}
O.~Kotlicki, J.~Scheuer, and M.~Shahriar, ``Theoretical study on brillouin
  fiber laser sensor based on white light cavity,'' {\em Opt. Express} {\bf
  20}, 28234--28248  (2012).

\bibitem{Phillips2013}
N.~B. Phillips, I.~Novikova, E.~E. Mikhailov, D.~Budker, and S.~M. Rochester,
  ``Controllable steep dispersion with gain in a four-level \emph{N}-scheme
  with four-wave mixing,'' {\em J. Mod. Opt.} {\bf 60}, 64--72  (2013).

\bibitem{SPIEProc2013}
I.~Novikova, E.~E. Mikhailov, L.~Stagg, D.~Budker, and S.~M. Rochester,
  ``Tunable lossless slow and fast light in a four-level \emph{N}-system,''
  {\em Proc. SPIE: Advances in Slow and Fast Light} {\bf 8636}, 86360C  (2013).

\bibitem{lukin03rmp}
M.~D. Lukin, ``Colloquium: {T}rapping and manipulating photon states in atomic
  ensembles,'' {\em Rev. Mod. Phys.} {\bf 75}(2), 457  (2003).

\bibitem{lukinPRLFWM99}
M.~D. Lukin, A.~B. Matsko, M.~Fleischhauer, and M.~O. Scully, ``Quantum noise
  and correlations in resonantly enhanced wave mixing based on atomic
  coherence,'' {\em Phys. Rev. Lett.} {\bf 82}(9), 1847  (1999).

\bibitem{Mikhailov2002}
E.~E. Mikhailov, Y.~V. Rostovtsev, and G.~R. Welch, ``Experimental study of
  {S}tokes fields linewidth in resonant four-wave mixing in {R}b vapour,'' {\em
  J. Mod. Opt.} {\bf 49}(14), 2535--2542  (2002).

\bibitem{phillipsJMO09}
N.~B. Phillips, A.~V. Gorshkov, and I.~Novikova, ``Slow light propagation and
  amplification via electromagnetically induced transparency and four-wave
  mixing in an optically dense atomic vapor,'' {\em J. Mod. Opt.} {\bf 56}(18),
  1916--1925  (2009).

\bibitem{LukinPhysRevA.79.013806}
T.~Hong, A.~V. Gorshkov, D.~Patterson, A.~S. Zibrov, J.~M. Doyle, M.~D. Lukin,
  and M.~G. Prentiss, ``Realization of coherent optically dense media via
  buffer-gas cooling,'' {\em Phys. Rev. A} {\bf 79}, 013806  (2009).

\bibitem{PhillipsPhysRevA.83.063823}
N.~B. Phillips, A.~V. Gorshkov, and I.~Novikova, ``Light storage in an
  optically thick atomic ensemble under conditions of electromagnetically
  induced transparency and four-wave mixing,'' {\em Phys. Rev. A} {\bf 83},
  063823  (2011).

\bibitem{FleischhakerPRA08}
R.~Fleischhaker and J.~Evers, ``Four-wave mixing enhanced white-light cavity,''
  {\em Phys. Rev. A} {\bf 78}, 051802  (2008).

\bibitem{AbiSalloum2009}
T.~Abi-Salloum, S.~Meiselman, J.~P. Davis, and F.~A. Narducci, ``Four-level
  `{N}-scheme' in bare and quasi-dressed states pictures,'' {\em J. Mod. Opt.}
  {\bf 56}(18), 1926--1932  (2009).

\bibitem{LettFWM12}
R.~T. Glasser, U.~Vogl, and P.~D. Lett, ``Stimulated generation of superluminal
  light pulses via four-wave mixing,''  (2012).
\newblock arXiv:1204.0810.

\bibitem{Siegman1986}
A.~E. Siegman, {\em Lasers}, University Science Books  (1986).

\bibitem{mikhailov04}
E.~E. Mikhailov, V.~A. Sautenkov, I.~Novikova, and G.~R. Welch, ``Large
  negative and positive delay of optical pulses in coherently prepared dense
  {R}b vapor with buffer gas,'' {\em Phys. Rev. A} {\bf 69}(6), 063808  (2004).

\end{thebibliography}

\end{spacing}
\end{document}